\pdfoutput=1
\documentclass[sigconf]{acmart}
\twocolumn

\renewcommand\footnotetextcopyrightpermission[1]{} % removes footnote with conference information in first column
\usepackage{multirow}
\usepackage{booktabs} % For formal tables

\usepackage[norelsize]{algorithm2e} % For algorithms

\SetAlFnt{\small}
\SetAlCapFnt{\small}
\SetAlCapNameFnt{\small}
\SetAlCapHSkip{0pt}
\IncMargin{-\parindent}

\begin{document}
\newcommand{\todo}[1]{\textcolor{red}{}}
\newcommand{\hetu}[1]{\textcolor{blue}{}}
\newcommand{\aaron}[1]{\textcolor{blue}{}}

% Title portion
\title{Match-Tensor: a Deep Relevance Model for Search}
\author{Aaron Jaech}
%\author{\em Author information deleted for blind review}
\email{ajaech@uw.edu}
\affiliation{%
  \institution{University of Washington}
  \streetaddress{185 E Stevens Way NE}
  \city{Seattle}
  \state{WA}
  \postcode{98195}
  \country{USA}}
\author{Hetunandan Kamisetty}
\email{hetu@fb.com}
\affiliation{%
  \institution{Facebook Inc}
  \streetaddress{1101 Dexter Ave N}
  \city{Seattle}
  \state{WA}
  \postcode{98109}
  \country{USA}}
\author{Eric Ringger}
\email{eringger@fb.com}
\affiliation{%
  \institution{Facebook Inc}
  \streetaddress{1101 Dexter Ave N}
  \city{Seattle}
  \state{WA}
  \postcode{98109}
  \country{USA}}
\author{Charlie Clarke}
%\author{\em Author information deleted for blind review}
\email{claclark@gmail.com}
\affiliation{%
  \institution{Facebook Inc}
  \streetaddress{One Hacker Way}
  \city{Menlo Park}
  \state{CA}
  \postcode{94025}
  \country{USA}}
\begin{abstract}
The application of Deep Neural Networks for ranking in search engines may obviate the need for the extensive feature engineering common to current learning-to-rank methods. However,  we show that combining simple relevance matching features like BM25 with existing Deep Neural Net models often substantially improves the accuracy of these models, indicating that they do not capture essential local relevance matching signals. We describe a novel deep Recurrent Neural Net-based model that we call Match-Tensor. The architecture of the Match-Tensor model simultaneously accounts for both local relevance matching and global topicality signals allowing for a rich interplay between them when computing the relevance of a document to a query.  On a large held-out test set consisting of social media documents, we demonstrate not only that Match-Tensor outperforms BM25 and other classes of DNNs but also that it largely subsumes signals present in these models.
\end{abstract}

% \ccsdesc[500]{Information systems~Retrieval models and ranking}
% \ccsdesc[300]{Computing methodologies~Machine learning}


% We no longer use \terms command
%\terms{Adhoc Retrieval, Search, Neural Networks}

%\thanks{ Author's addresses: Aaron Jaech, Electrical Engineering Department, University of Washington {and} J. A. Stankovic, Computer Science
%  Department, University of Virginia; T. Yan, Eaton Innovation Center;
%  T. He, Computer Science Department, University of Minnesota; C.
%  Huang, Google; T. F. Abdelzaher, (Current address) NASA Ames
%  Research Center, Moffett Field, California 94035.}

\maketitle

\section{Introduction}

% Figure
\begin{figure*}[ht]
\includegraphics[width=0.8\textwidth]{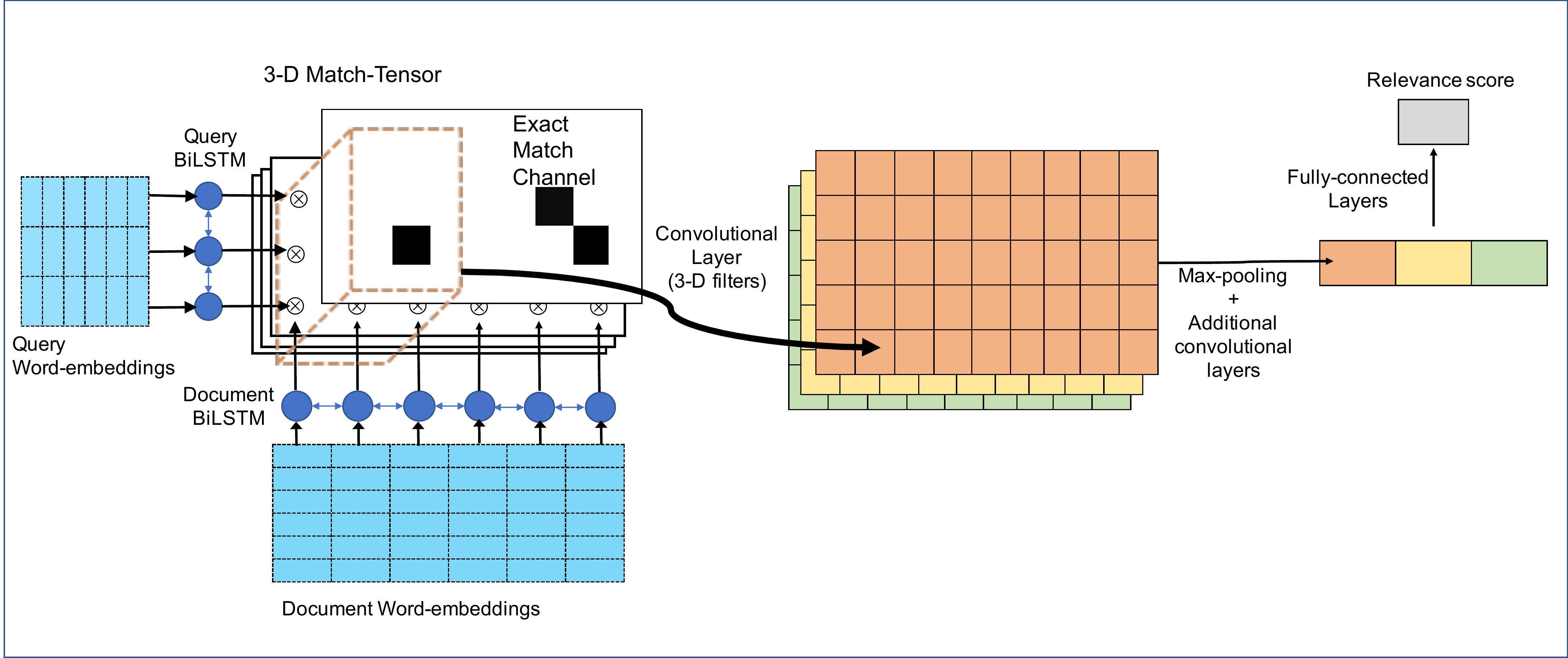}
\caption{Match-Tensor architecture. Word embeddings are mapped to bi-LSTM state separately on the query and document. The transformed bi-LSTM state is combined by way of point-wise product for each query term and each document term to produce multiple match channels.  The match channels are joined by an exact-match channel to produce the 3-D Match-Tensor.  2-D convolution (with 3-D filters) maps the Match-Tensor to a probability of relevance for the query and the document.}
\label{fig:architecture}
\end{figure*}

Machine learning models have traditionally relied on a set of manually defined features whose weights are trained in order to optimize an objective function on a set of labeled examples.
The learning-to-rank models in widespread use for search
may require hundreds of such features to be generated for each query/document
pair~\cite{Macdonald:2013:WHL:2559123.2559126}.
In particular, the approach requires features that capture the presence of query terms in a given document,
the proximity and location of those terms in the document, document quality,
document type, and many other important aspects of the query, of the document,
and of the relationship between them.
In many rankers, the feature set includes classic information retrieval models,
such as BM25~\cite{bm25},
and term dependence models~\cite{Metzler:2005:MRF:1076034.1076115}.
These models in turn combine more basic query and document features in an
attempt to capture salient aspects of relevance.

Generation of these features requires a substantial software engineering effort,
and the addition of a single new feature may require the creation and testing of
new and specialized code to compute it.
More importantly, the scope of the feature set is limited by the imagination
of its engineers.
If the feature set fails to capture some aspect of relevance,
that aspect is lost to the ranker.

Recently Deep Neural Networks (DNNs) have replaced this traditional machine
learning paradigm in many applications by directly learning a model and its
features from the inputs, obviating the need for manually specified features.
Motivated by success in machine translation and other related NLP tasks,
DNNs are now being applied to ranking in search \cite{dssm,c_dssm,shen2014latent,mitra2015exploring,nalisnick2016improving,kenter2015short,yang2016anmm} \todo{check if we should add others here?}.
These efforts share a goal of reducing the need for traditional feature
engineering, and even if DNNs cannot completely eliminate feature engineering,
they may capture aspects of relevance beyond its scope.

Unfortunately, while previous efforts have demonstrated that these approaches
can capture novel aspects of relevance, they can also fail to capture the most salient aspect of the search task,
namely local relevance matching~\cite{croft_relevance_cikm16}.
Indeed, some recent efforts to develop completely new DNN architectures
tailored to search still show lower accuracy on TREC datasets than the classic information retrieval models, which are used as features in traditional learning-to-rank approaches~\cite{matchpyramid1,matchpyramid_study}.
While this failure may possibly be due to the lack of sufficient training data, which is critical to the success of DNNs, the failure may also reflect a more fundamental gap in these architectures.
\todo{Specifically address the volume of training data with a learning curve. The question of data volume is one that we can speak to better than most groups, so we should do so.}

We develop a new deep neural net architecture tailored for the search task involving social media using a method we refer to as \emph{Match-Tensor}.
The model architecture is simultaneously simple and expressive and is schematically illustrated in Fig.~\ref{fig:architecture}:
a pair of sequence models computes representations of a query and a given document accounting for both local and distributed context; these representations are used in matching each pair of query and document words along several dimensions and stored as a 3-D tensor (hence, the term Match-Tensor).
A convolutional network then translates this tensor to a relevance score.
Using a large volume of social media search data, we train a model of this type and analyze the performance, demonstrating that it is also substantially more accurate than other current model-classes.
We also demonstrate that this approach largely \emph{subsumes} other models such as BM25\cite{bm25} and SSM\cite{dssm,c_dssm}:
the Match-Tensor model alone performs nearly as well as meta-models trained with both Match-Tensor and these features as input.

\section {Background}

Like most learning-to-rank methods, we wish to learn a function $\Phi(q,d)$
that computes a score reflecting the relevance of document $d$ for query $q$.
In the traditional feature engineering approach, we would start with a set of hand-crafted features $F$
that capture various aspects of relevance matching, combine them in a single model $M$ -- such as logistic
regression or boosted decision trees -- and train the model
using a learning-to-rank approach~\cite{letor_cao2007,lambdamart}
to predict the labels on training data:
\[
\Phi(q,d) = M(F(q,d)) \]
The features employed in this approach may be as simple as binary query term presence in the document
or as complex as separately trained classification or ranking sub-models.
Furthermore, it is standard to include classic information retrieval models in this
feature set, particularly BM25~\cite{bm25}.
Liu~\cite{ltrliu} provides a thorough overview of traditional learning-to-rank methods for search.
Macdonald et al.~\cite{Macdonald:2013:WHL:2559123.2559126} cover many of the engineering issues associated with deploying learning-to-rank in a search engine.

The advent of Deep Neural Networks has led to the development of an exciting alternative:
In this approach, a single learning procedure is used to learn both features and a model \emph{simultaneously}.
Huang et al.~\cite{dssm} introduced the first Deep Neural Network architectures for Web search that operated on (query, title) pairs, using a so-called \emph{siamese} architecture~\cite{lecun1995convolutional}, in which two feed-forward networks $NN_Q$ and $NN_D$ map the query $q$ and the title of a given web document $d$, respectively, into fixed-length representations:
\[
\Phi(q,d) = \cos(NN_Q(q), NN_D(d)),\]
The final documents are then ranked by their similarity to the query in this space computed using cosine similarity.
The application of convolutional neural networks, in lieu of feed-forward-networks, by Shen et al.~\cite{c_dssm} marks the next notable advancement using the same siamese architecture.
The local connectivity of convolutional networks can allow for more accurate models, especially when it \todo{the connectivity?} mirrors the structure of the task at hand.

In parallel to these developments, there have been several advances in Deep Neural Networks, especially for modeling text.
While earlier approaches to DNNs for text used convolutional networks, more recent approaches have used Recurrent Neural Networks (RNNs), especially those based on Long Short-term Memory (LSTM) units~\cite{LSTM}.
Unlike convolutional networks, the units in recurrent networks maintain an internal state that is updated from word to word as a given text is processed, allowing for the network to capture sequential relations across a query or document.
A popular architecture for machine translation uses the so-called sequence-to-sequence paradigm in which the input text in the source language is ``encoded'' using an encoder network to produce a fixed-length representation (the RNN state)\cite{sutskever2014sequence}.
A ``decoder'' then begins with this representation and emits an output in the target language.
While the use of a fixed-length representation is similar to the architecture of Huang et al.~\cite{dssm} and Shen et al.~\cite{c_dssm}, the use of RNNs such as those based on LSTMs is critical to their performance.
Attention-based schemes build on this architecture by dynamically re-weighting (i.e., focusing attention) on various elements of the source representation during the decoding process, and they have demonstrated considerable improvements over their non-attention counterparts~\cite{bahdanau2014neural}.

The ``representation-based`` nature of siamese architectures has also been identified as a limitation in search~\cite{croft_relevance_cikm16} and has led to the development of alternate ``interaction-based'' architectures, in which the relationships between query and document are considered earlier.
In an approach called Match-Pyramid, Pang et al.~\cite{matchpyramid1} construct an interaction matrix between query and document terms, where each entry in the matrix denotes the strength of similarity between the corresponding terms.
A hierarchical convolutional model then operates on this single interaction matrix to compute the final score \todo{what is the difference between this and \cite{arc}-II?}.
A recent paper by Mitra et al.~\cite{mitra2016learning}, that
appeared while this manuscript was being prepared, uses a ``duet'' architecture in which two separate networks (one ``representation''-based and the other ``interaction''-based) are combined to simultaneously account for local and distributed measures of similarity.
The key idea in this method is to use an exact-match matrix followed by convolutional layers on the ``interaction'' half of the network in addition to a siamese architecture.
A crucial limitation of such an approach to modeling interactions is that all tokens in the query are given equal importance: the interaction model can therefore not distinguish between query terms that are important and those that are not \cite{salton1988term}.

To address some of these shortcomings, we developed a new architecture that we refer to as \emph{Match-Tensor}.
Our thesis is that no single interaction matrix can capture all aspects of the matching problem.
Unlike previous interaction-based approaches, which use a single representation of the interaction matrix and cannot distinguish between different words that have the same pattern of matching, the Match-Tensor model architecture simultaneously considers similarity along several dimensions during the matching process.
This approach allows for a rich interplay between different dimensions in subsequent stages of the architecture in order to determine the relevance of a document to the query.

\section{Match-Tensor Architecture} \label{sec:model}

\begin{figure}[ht]
  \includegraphics[height=0.7\textheight]{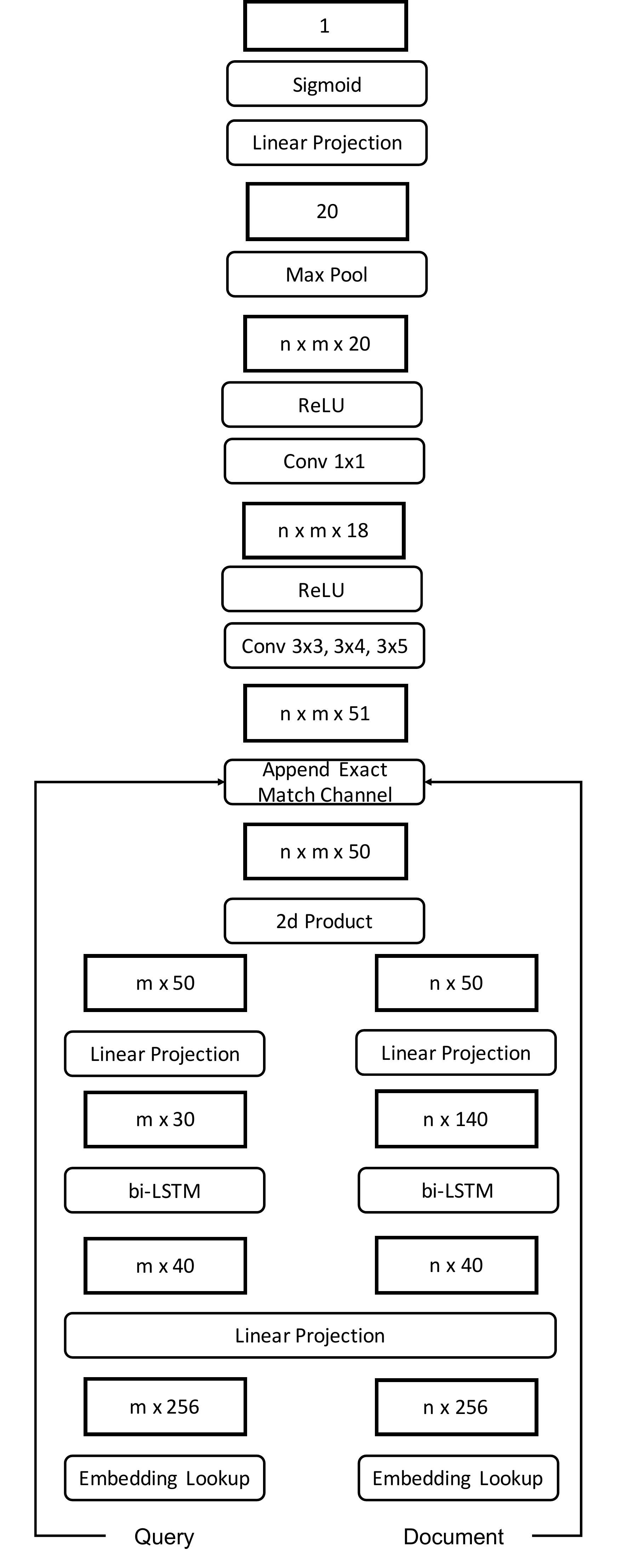}
\caption{Match-Tensor Architecture in Detail. Individual elements of the architecture are shown in rounded boxes, while tensor state is shown in the rectangular boxes.  Query and document representations are combined in the match-tensor, from which a probability of relevance is finally computed.}
\label{fig:model_detail}
\end{figure}

The central thesis of the Match-Tensor architecture is that it is vital to incorporate multiple notions of similarity, capturing both immediate and larger contexts in a given document when computing the relevance of that document to a query.
This objective is achieved by a three-dimensional tensor, in which one dimension corresponds to terms in the query, one dimension for the terms in the document, and a third dimension for various match channels.
Each match channel contains a distinct estimate of the match similarity between the query and document, hence the name ``Match-Tensor''.
The tensor is computed using the output of a neural network operating on word-embeddings and is supplemented with an exact-match channel that operates directly on the tokens;
a downstream neural network is then employed to determine the relevance of the document to the query using the tensor.
The entire network is trained end-to-end with a discriminative objective.
Thus, the manner in which these multiple notions of similarity are combined to produce the final relevance score is deferred until after all channels are computed.
Fig.\ref{fig:model_detail} illustrates the final model in detail.  In what follows, we motivate and introduce each element of the architecture.

\todo{in figure 2, label the actual box that is the ``match-tensor'' with that name}

% Figure

\subsection{Input to the Match-Tensor Layer}

To begin, a word-embedding lookup layer converts query and document terms into separate sequences of word-embeddings.
The word embedding table is itself computed offline from a large corpus of social media documents using the word2vec package~\cite{word2vec} in an unsupervised manner and are held fixed during the training of the Match-Tensor network. In our implementation, word-embeddings are 256-dimensional vectors of floating point numbers.
The word embeddings are then passed through a linear projection layer to a reduced $l$-dimensional space (here $l = 40$); the same linear projection matrix is applied to both the query and the document word vectors.
This linear projection allows the size of the embeddings to be varied and tuned as a hyperparameter without relearning the embeddings from scratch each time.
Two Recurrent Neural Networks, specifically bi-directional LSTMs (i.e, bi-LSTMs)~\cite{LSTM,biLSTMs}, then encode the query (respectively document) word-embedding sequence into a sequence of LSTM states.
The bi-LSTM states capture separate representations in vector form of the query and the document, respectively, that reflects their sequential structure, looking beyond the granularity of a word to phrases of arbitrary size.
During hyperparameter tuning, we allowed the models to use a linear projection layer inside the bi-LSTM recurrent connection, as defined in Sak et al.~\cite{sak2014long}, but the best models did not make use of it.
There is a separate linear projection after the bi-LSTM to establish the same number $k$ of dimensions in the representation of query and document (in our case, $k = 50$).
Thus, at the end, each token in the query and the document is represented as a $k$-dimensional vector.

\subsection{Match-Tensor Layer}

For $m$ words in the query and $n$ words in the document, the actual match-tensor -- from which the architecture inherits its name -- is an $m \times n \times k+1$ tensor, where $k+1$ is the number of channels in the match-tensor (specifically, 51 in the detailed figure).
Each of the $k+1$ channels is computed from a distinct representation of the query and document: all but one of the channels are computed using the element-wise product of the corresponding bi-LSTM states of the query and document (after applying the subsequent projection).
Including each dimension as a separate layer instead of collapsing them into a single layer allows the model to include state-specific (approximately: term-specific) signals in the matching process and to weigh matching different terms according to their importance.
While this approach captures most of the key signals, one omission of the first $k$ layers is their inability to properly represent out-of-vocabulary tokens in the query or document, beginning in the initial word embedding lookup.
To compensate for this problem, the initial embedding lookup includes an out-of-vocabulary vector, and the model appends an extra exact-match channel in the match-tensor (hence, $k+1$ total channels) such that the entry at position $i, j$ of this channel is set to $\alpha$ if word $i$ in the query is an exact match to word $j$ in the document and zero otherwise.
This exact-match channel is critical for capturing local textual match.
The value of $\alpha$ is learned via back-propagation along with the rest of the model.

\subsection{From Match-Tensor to Score}

The secondary neural network begins with the match-tensor and applies a convolutional layer: the match-tensor is convolved cross the full depth ($k+1$) of the tensor with three sets of filters, each having a width of three query words and a height of three, four, or five document words.
As illustrated in Fig.~\ref{fig:architecture}, these 3-D convolution filters enable the model to learn interactions among the representations in ways that would be very difficult to anticipate as a feature engineer, lending expressive strength to the model architecture.
The model applies a ReLU (rectified linear unit) function to the output of these convolutions and then convolves that output with a set of $1 \times 1$ filters.
The ReLU activation function was chosen because it has been shown to be effective for convolutional neural
networks in computer vision \cite{jarrett2009best}.
Finally, the model applies 2-D max-pooling to coalesce the peaks from the ReLU into a
single fixed size vector.
This is fed into a fully-connected layer (labeled as ``linear projection'' in the diagram) and through a sigmoid
to produce a single probability of relevance on the output of the model.

\section{Additional Related Work}

%\todo{there is significant redundancy between the first part of this section and the background section.  Charlie says: don't worry about it for now}
The work presented in this paper follows a rich and rapidly growing body of work that use Deep Neural Networks in Search \cite{dssm,c_dssm,shen2014latent,craswellneu,kenter2015short,mitra2015exploring,yang2016anmm,nalisnick2016improving}. Our work is closest to the so-called Match Pyramid models of Pang et al.~\cite{matchpyramid1,matchpyramid_study}: Match Pyramid models construct a single match matrix and then use a convolutional network on top of it (hence ``pyramid'') to compute a relevance score. Unlike them, the Match-Tensor architecture developed in this work simultaneously considers multiple channels during the matching process allowing for a rich interplay between the different channels in determining the relevance of a document to a query. Match Pyramid models are unable to distinguish between different words having the same match pattern. Guo et al.~\cite{croft_relevance_cikm16} developed a neural-network based model (DRMM) that uses matching histograms and term-gating. They report that this model is more accurate than BM25 and other alternatives on standard TREC test collections (Robust-04 and ClueWeb-09-Cat-B) however Mitra et. al \cite{mitra2016learning} report that a model incorporating an exact-match channel with a representation based ``distributed'' model outperforms DRMM on a larger collection of web-search queries.

Several other models that use word-embeddings have been proposed for various aspects of Search including  \citet{diaz2016query} who use them in query expansions, \citet{ganguly2015word} who use them in smoothing language models, \citet{nalisnick2016improving} who propose dual embeddings and Grbovic et al.~\cite{grbovic2015search, grbovic2015context} who use them in sponsored search. Cohen et al.~\cite{croft_adaptability} have also studied the utility of DNNs for several IR tasks.

%Zheng and Callan [42] use term embeddings
%as evidence for term weighting, learning regression models to
%%optimize weighting in a language modeling and a BM25 retrieval
%model. Ganguly et al. [8] used term embeddings for smoothing
%in the language modeling approach of information retrieval. Nalisnick
%et al. [28] used dual embeddings, one for document terms
%and one for query terms, then ranked according to the all-pairs
%similarity between vectors. Diaz et al. [6] used term embeddings
%to generate query expansion candidates in the language modeling
%retrieval framework, also finding better performance when training
%a specialized term embedding.

\section{Methodology}

\subsection{Data}

We base our experiments on a collection of approximately 1.6 million (query,document,label) triplets, collected on a major social
media site between between 2016-01-01 and 2016-06-01.
Each document is a publicly viewable social media post, which might
include videos, photos, and links, as well as text, but
for the experiments reported in this paper, we consider only the text.
Labels indicate the relevance level of the document with respect to the
query.
For these experiments, we use three levels of relevance:
``VITAL'', ``RELEVANT'', and ``NONRELEVANT''.
We split this dataset (by query) into 3 parts: train, validation, and test, so that each query string appears in exactly one of the partitions. We provide details of the partitioning in table~\ref{table:dataset}.
The training and validation sets were used to train the models and perform hyper-parameter sweeps. The test set was used only for evaluation, at the end of this process.

\begin{table}[]
\centering
\caption{Details of Dataset. Our corpus consists of approximately 1.6 M (query,document,label) triplets that we partitioned into train,validation and test. Each query string and the corresponding results appear in exactly one of the partitions. The test partition was used only during the final evaluation.}
\label{table:dataset}
\begin{tabular}{lrrr}
 & \textbf{\begin{tabular}[c]{@{}l@{}}Unique\\ Queries\end{tabular}} & \textbf{Results} & \textbf{\begin{tabular}[c]{@{}l@{}}Average\\ Results/Query\end{tabular}} \\ \hline
Train& 59,457 & 1,032,325 & 17.36\\
Validation & 3,975 & 69,005 & 17.36 \\
Test & 35,773 & 615,242 & 17.20\\
\end{tabular}
\end{table}

\subsection{Implementation Details}

We implemented our Match-Tensor neural net model using TensorFlow~\cite{tensorflow2015-whitepaper}.
We used pre-trained 256-dimensional phrase embeddings using the word2vec package~\cite{mikolov2013efficient} on a large corpus of documents with a vocabulary size of around 2 million tokens containing unigrams and selected bigrams.
Out-of-vocabulary words are mapped to a special token.
Queries are truncated to a maximum of eight words in length, whereas documents are truncated to a maximum length of 200 words.
Both the query and the documents are then preprocessed by lowercasing and applying a simple tokenizer to split words and
remove punctuation.
Since social media documents are structured into separate fields (e.g. the title, author, and body), we added special tokens for demarcating the boundaries between fields and for the start and end of a document.
The embeddings for these special boundary tokens are randomly initialized and kept fixed during training.
We used dropout as a regularizer on the non-recurrent connections of all bi-LSTMs.
We employed the Adam optimizer for gradient descent~\cite{kingma2014adam} with a learning rate of 0.001 and mini-batch size of 200.
Hyperparameter settings are shown in \ref{table:hyperparameters}.

To investigate the importance of each of the major components in the model we also experimented with alternative choices for these components and alternate architectures, tailoring them to social media documents. While some architectures have been suggested for short text common in some social media\cite{lu2013deep,matchpyramid1}, studies indicate that they do not beat baseline models such as BM25\cite{matchpyramid_study}. In contrast, both early models \cite{c_dssm,dssm} and recent developments by Mitra et.al that has shown strong performance \cite{mitra2015exploring} have been designed for web-search and are not directly usable in our setting. We therefore adapted these model architectures from web search for our setting. The details are discussed in the following sections.

\subsection{Semantic Similarity Model(SSM)}

We constructed an model using the siamese network architecture based on the semantic similarity models (SSM) appearing in other work~\cite{dssm,c_dssm,palangi2014semantic}.
In this SSM model shown in detail in Figure \ref{fig:model_detail_ssm}, we construct a query embedding by concatenating the last output from each of the forward and backward directions of the query bi-LSTM and a document embedding by max-pooling over the output bi-LSTM states across the entire document.
Max-pooling is used for the document because the documents can be much longer than the query and it is harder for the bi-LSTM to propagate the relevant information all the way to the end of the document~\cite{lai2015recurrent}.
These fixed-length document and query embeddings are then passed throw  linear projections before computing a dot-product between them which is then used to compute the final score.
The model parameters and hyper-parameters were optimized on the same dataset as the Match-Tensor model.

\subsection{Match-Tensor(Exact-only)+SSM}

In recent work that appeared while this manuscript was being prepared, Mitra et al \cite{mitra2016learning} show that a combination of local and distributed matching can outperform other models for web-search. Since several details of their model are specific to the structure of web documents, we constructed an model that has similar characteristics for our settings by combining a single channel exact-match only Match-Tensor component with an SSM component into a single model.
This is done by concatenating the output from the last layer of Match-Tensor filters with the hidden layer of the SSM comparison network as shown in Figure \ref{fig:model_detail_match_plus_ssm}.
The Match-Tensor and SSM components share parameters for the word embedding and LSTM portion of the model.

\subsection{Match-Tensor+SSM}

We also compare the effect of using all the channels in the Match-Tensor architecture in conjunction with the SSM architecture. This model is shown in Figure \ref{fig:model_detail_match_plus_ssm}. The only difference in architecture between this model and the previous (exact-match only channel) model is the number of channels in the tensor layer: the former has one channel while this model has $k+1$ like the Match-Tensor model.
\subsection{bi-LSTMs vs CNNs}

We compared all three model architectures listed against similar ones that uses convolutional layers in-place of bi-LSTMs.
We used a mix of width 1 and width 3 convolutional filters. Compared to the bi-LSTMs, that can incorporate information over a wide token span, the representations produced by the convolutional filters
only look at trigrams (when the width is 3) but are computationally cheaper.

\subsection{Attention Pooling}

To improve on the query-agnostic pooling schemes of SSM, we also implemented an attention pooling mechanism for the document embeddings as an
alternative to max pooling.
The hypothesis here is that information from the
query is important in deciding in how to summarize the document.
The attention
pooling model learns a ReLU activated transformation from the query embedding
and each output from the document bi-LSTM.
Attention weights are determined by taking the dot product between these vectors and normalized using the Softmax function.
The attention-pooled document embedding is the weighted combination of the bi-LSTM outputs.
We note that our use of attention is different from that of Palangi et al.~\cite{zhai2016deepintent} where attention-based pooling was used in a query-agnostic manner.
In our preliminary experiments, using attention-based pooling did not result in improved results compared to a max pooling baseline so we did not pursue it further.

\subsection{Ensemble models}
\aaron{is this section redundant given section 6.6?}
Comparing different model architectures using absolute metrics can yield insight into the relative importance of the types of signals for the task at hand.
However, one model might outperform another model without capturing the signal in the latter model.
Consequently, to test if one model \emph{subsumes} another, we train additional ensemble models that use
the scores of both models. We measure the accuracy of the ensemble models in addition to the individual models.

\section{Results}

\subsection{Model selection}
We optimized hyperparameters based on a random grid search on the validation set for each model architecture studied, selecting the one model with the best score out of 200 runs.
For each model architecture we next evaluated the single best model on the test set.
Table~\ref{table:hyperparameters} reports these hyperparamters for the three main model architectures we studied.
Critically, we note that the final Match-Tensor model has \emph{fewer} parameters than the final SSM model.

\begin{table*}[htb]
\centering
\caption{Hyperparameter settings for each model architecture. These were selected by performing a random grid search on the validation set and selecting the best setting of 200 runs.}
\label{table:hyperparameters}
\begin{tabular}{lccc}
\multicolumn{1}{c}{}        & \textbf{SSM} & \textbf{Match-Tensor} & \textbf{Match-Tensor+SSM} \\ \hline
Word Embedding Projection   & 50           & 40                    & 50                        \\
Doc. bi-LSTM Dim.           & 120          & 70                    & 95                        \\
Query bi-LSTM Dim.          & 32           & 15                    & 15                        \\
Comparison Net Hidden Layer & 50           & 50                    & 55                        \\
Match-Tensor Size           & --           & 40                    & 35                        \\
Match Filters 1st Layer     & --           & 18                    & 18                        \\
Match Filters 2nd Layer     & --           & 20                    & 30                        \\
Training Epochs             & 4.25         & 4.5                   & 3.25                      \\
Total Parameters            & 216K         & 104K                  & 160K
\end{tabular}
\end{table*}

\subsection{Sensitivity to Training Size}

\begin{figure}
\includegraphics[width=0.45\textwidth]{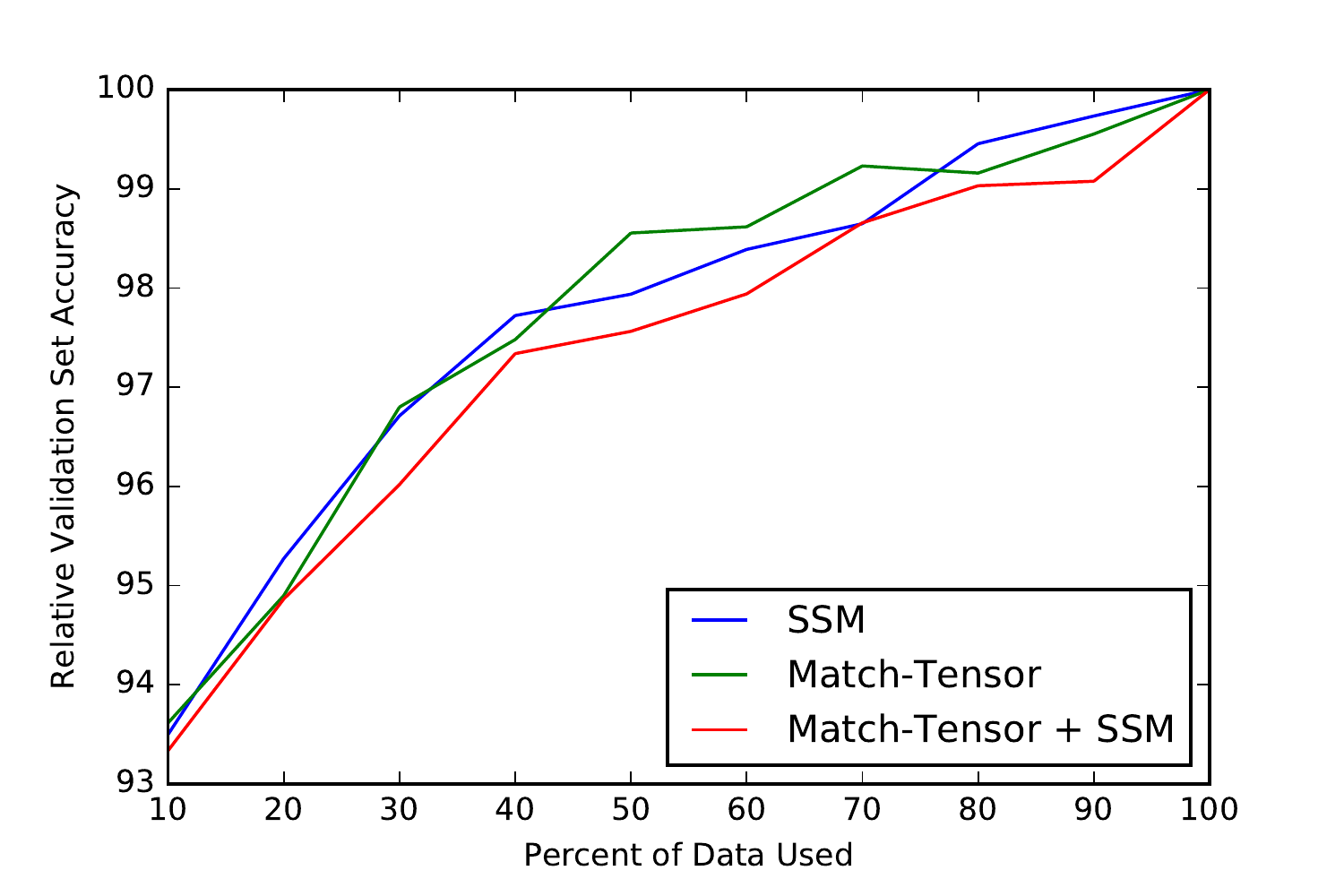}
\caption{Test-accuracy (AUC of ROC curve) as a function of size of training set for each model, relative to test-accuracy of model trained with all data. In each case, we fixed the hyperparameter settings to the values selected in Table \ref{table:hyperparameters}.}
\label{fig:training_size}
\end{figure}

To evaluate the sensitivity of the model performance to the amount of training data, for each of the NN architectures we sub-sampled the training set, retrained models (keeping the hyper-parameters fixed), and computed the test-loss.  Figure~\ref{fig:training_size} shows the test loss of each model as a function of \emph{its} final accuracy. Not surprisingly, each architecture we consider here benefits from the availability of large training sets, and the accuracy improves substantially as we increase the size of the training set.
However, the relative comparisons between the model architectures appear to be reasonably robust to training data size.

\subsection{Performance of Neural Models}

\begin{figure*}
\includegraphics[width=0.6\textwidth]{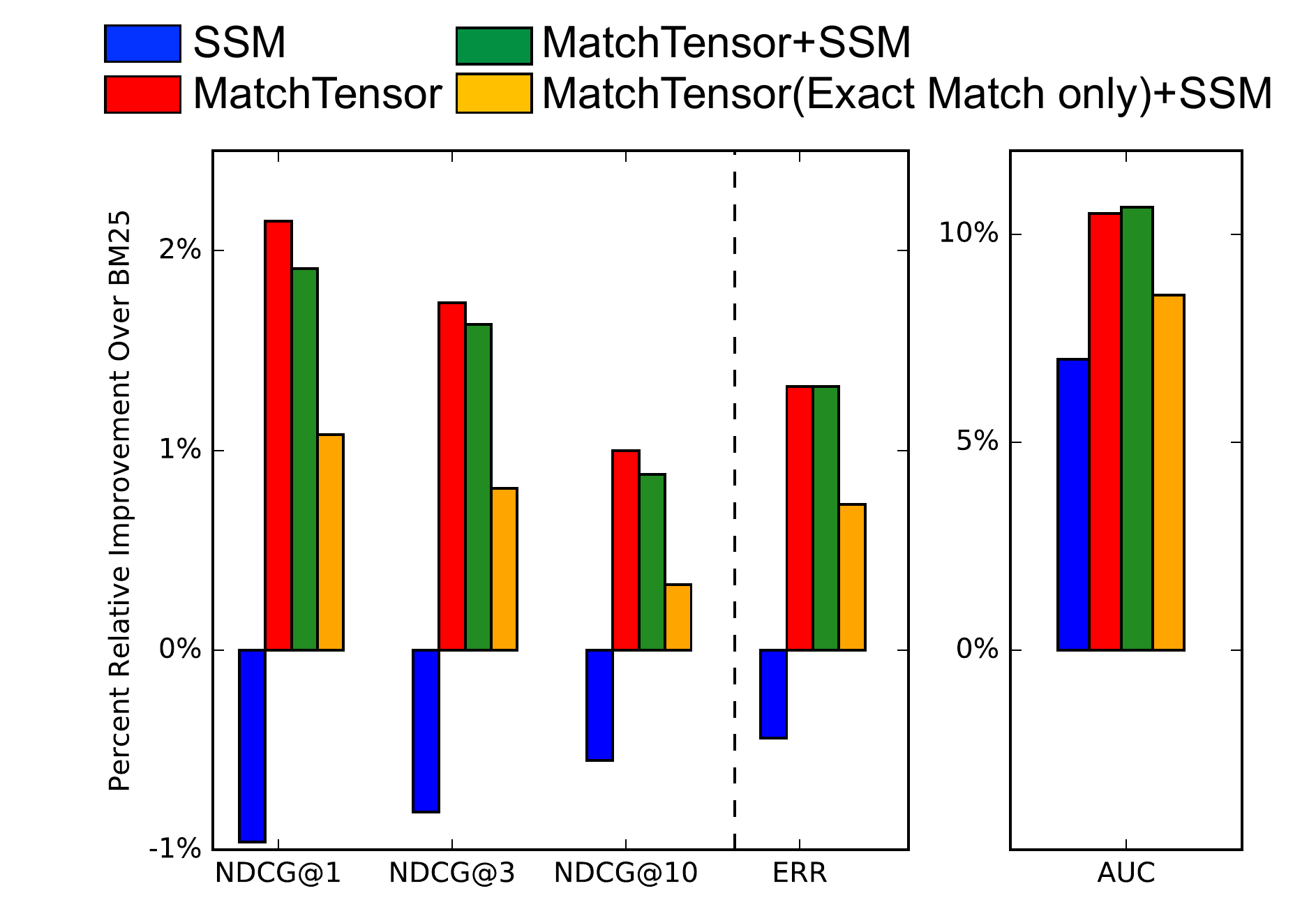}
\caption{Accuracy of different model architectures relative to BM25 (in percentage).}
\label{fig:irmetrics}
\end{figure*}

\begin{figure}[h]
  \includegraphics[height=0.7\textheight]{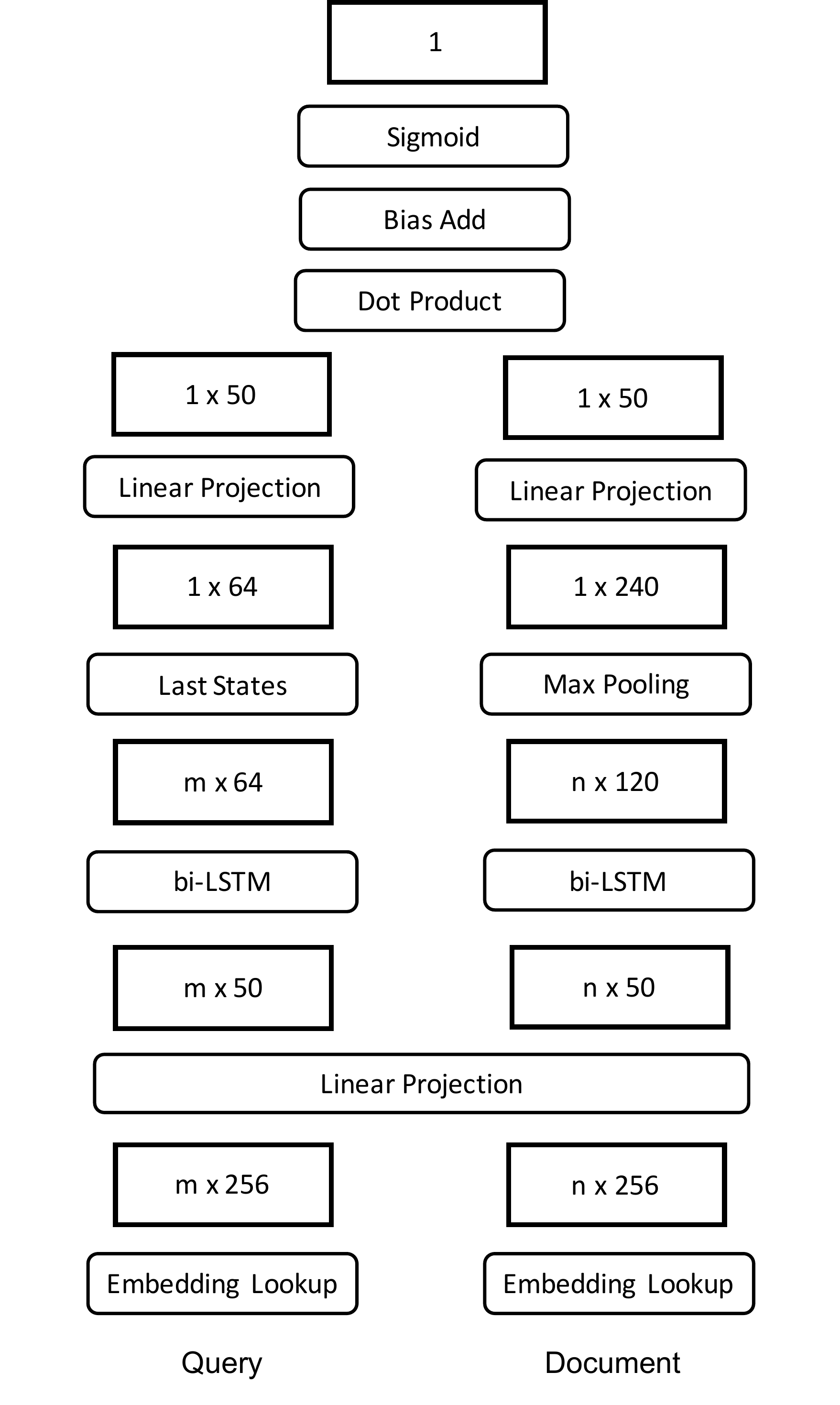}
\caption{Architecture of SSM model.  Individual architectural elements shown in rounded boxes; tensor state shown in rectangular boxes.  Query and document representations are combined via dot product to compute a probability of relevance.}
\label{fig:model_detail_ssm}
\end{figure}

\begin{figure}[ht]
  \includegraphics[height=0.7\textheight]{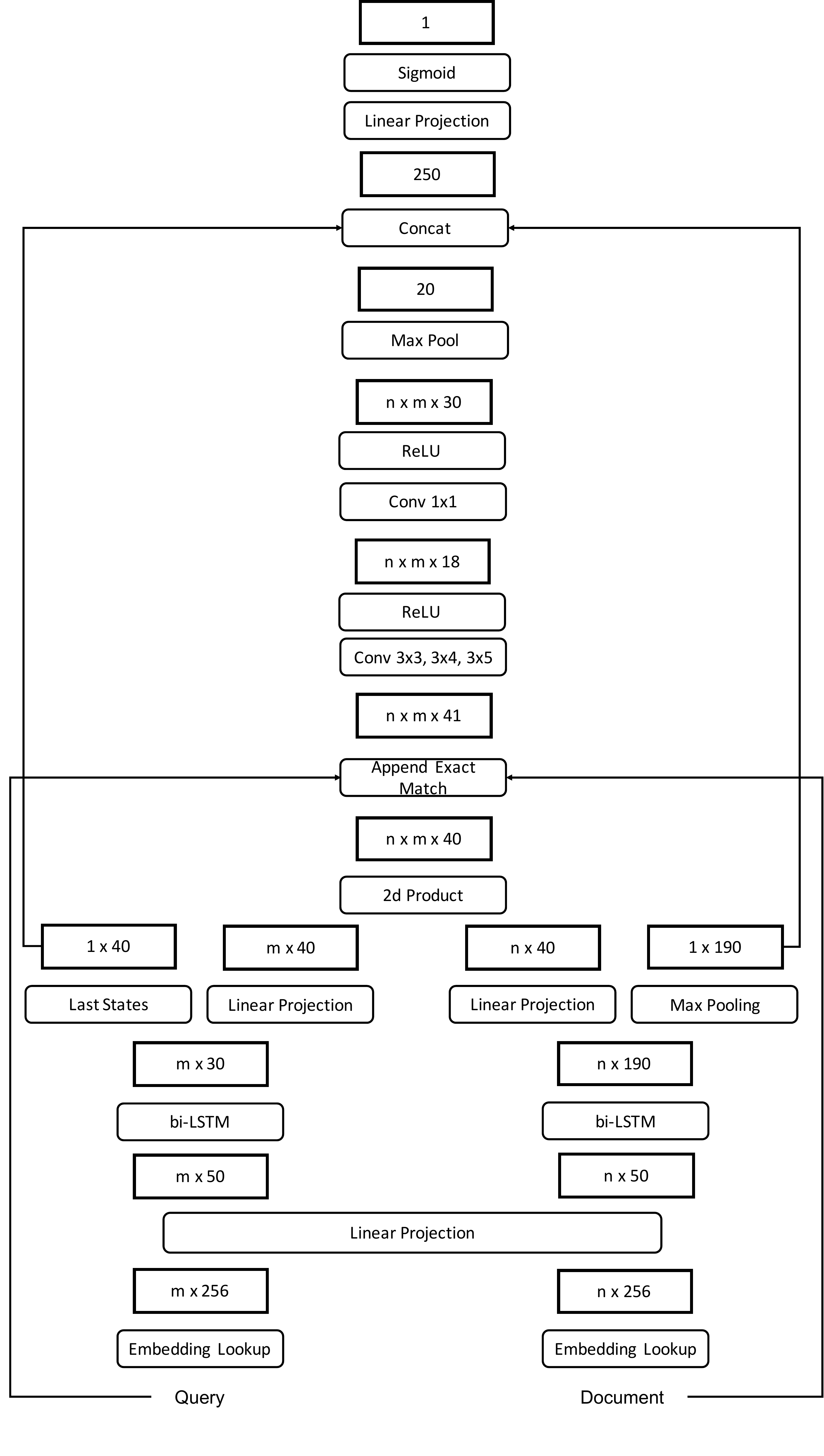}
\caption{Architecture of Match-Tensor+SSM model}
\label{fig:model_detail_match_plus_ssm}
\end{figure}

Figure~\ref{fig:irmetrics} summarizes the performance of the various neural model architectures relative to a BM25 baseline.
The figure reports NDCG at various levels as well as Expected Reciprocal Rank (ERR)~\cite{err},
with all measures computed using the three relevance grades.
Overall, the Match-Tensor model (with bi-LSTMs) is the most accurate individual model, with an 11\% improvement in AUC of the ROC curve (right panel of the figure) over a BM25 baseline and smaller but consistent improvements on NDCG and ERR. We note that while the relative ordering of models appears to be robust to variations in the test set, the values of these relative improvements appear to be sensitive to the composition of the test set: relative improvements when restricted to a subset of the test-set that are ``hard'' (at most half the available results are relevant) are much larger.

The SSM architecture had lower NDCG than the BM25 baselines.
This result is consistent with \cite{croft_relevance_cikm16} and others who have highlighted the limitation of models that only match semantic representations in relevance-matching tasks.
Match-Tensor is not only more accurate in aggregate, it is also more accurate at every cutoff: the Precision of the model is higher than the others at all values of Recall.

\begin{table*}[tp]
\centering
\caption{Relative improvement for each model architecture when using bi-LSTMs over using CNNs}
\label{table:cnn_vs_lstm}
\begin{tabular}{r|c|ccc|c}
  \textbf{Model} & \textbf{AUC} & \textbf{NDCG@1} & \textbf{NDCG@3} & \textbf{NDCG@10} & \textbf{ERR}  \\
\hline
SSM                                   &2.02\% &1.10\% &0.83\% &0.67\% &0.74\%\\
Match-Tensor                          &2.87\% &1.18\% &1.04\% &0.66\% &0.58\%\\
Match-Tensor+SSM                      &2.07\% &0.59\% &0.57\% &0.33\% &0.43\%\\
Match-Tensor (Exact Only)+SSM         &1.57\% &0.12\% &0.00\% &0.11\% &0.15\%\\
\end{tabular}
\end{table*}

\subsection{bi-LSTMs vs CNNs for text representation}

We considered the use of convolutional neural networks to compute the text representations in the first stage of each model architecture in place of bi-LSTMs.
Table~\ref{table:cnn_vs_lstm} shows that across the four model architectures under consideration, using bi-LSTMs results in more accurate models than their CNN counterparts in terms of AUC, NDCG, and ERR.
\todo{For table 3: consider using BM25 as the baseline and then show bi-LSTM vs. CNN side-by-side in chart format.}
For AUC in particular, the relative gain in AUC from using bi-LSTMs is between two to three percent.
The fact that this increase holds for both SSM and Match-Tensor architecture variants suggests that the improvements are due to bi-LSTMs -- across the board -- providing more accurate representations at each position.
This outcome is not surprising given similar results in other language modeling tasks \todo{include some citations of bi-LSTM vs CNN. I (Aaron) spent 10 minutes looking for a citation comparing bi-LSTM against CNN and couldn't find anything.} and is consistent with gains in NDCG observed in~\cite{palangi2014semantic} in going from convolutional to bi-LSTM-based semantic similarity models.
\subsection{2-D Matching vs. SSM}

As we have seen, the Match-Tensor architecture outperforms the SSM architecture, often significantly.
Although both architectures are most accurate when using bi-LSTMs, the relative improvement when going from SSM to Match-Tensor is substantial.
This improvement holds even when using CNNs to represent the state at each query and document token: AUC goes up by 4\% when using bi-LSTMs and 3\% when using CNNs, suggesting that the improvement is a consequence of the underlying difference in architectures.
The superiority of Match-Tensor is not surprising given that the Match-Tensor architecture has a substantially more expressive matching function.
Furthermore, combining the Match-Tensor and SSM architectures gives no substantial gain in performance: as seen, small improvements in AUC are offset by small reduction in NDCG and ERR.
The absence of a difference for this hybrid architecture would suggest that the bi-LSTM representation at each position is already capturing global context sufficiently well to make the additional explicit per-document representation non-informative for this problem.

\subsection{Influence of the exact-match channel}

While we introduced the exact-match channel to account for out-of-vocabulary tokens where the bi-LSTM states might not be accurate, it is computed for all cases.
Not surprisingly, it is an important contributor to the accuracy of the final model.
However the interplay between all channels improves the accuracy of the model further: the relative NDCG at 1 goes up by 2\% with the bi-LSTM channels on compared to the exact-match alone model where the relative improvement is roughly 1\% i.e., roughly half. This approximate doubling in relative accuracy when moving from the single channel to the full Match-Tensor model is seen across all positions in NDCG and in ERR.

\subsection{Ensemble models}

To determine if a deep relevance model is indeed capturing all essential relevance matching signals, we trained ensemble models: boosted trees \cite{friedman2001greedy} that combine as inputs the neural model's output as a feature and a BM25 feature
using 5-fold cross-validation on the existing validation set. Neural Models that capture essential relevance matching signals better should show relatively small improvements when BM25 is added to the mix, compared to those that don't since a good model should already capture most of the signal in BM25.
As seen in Table \ref{table:results}, Match-Tensor shows the smallest relative increase when BM25 is added to the mix compared to all other alternatives.
An exact-match only Match-Tensor +SSM model also does better in this regard than SSM alone although again the full Match-Tensor model is substantially better by allowing for interplay among channels even without having an explicit SSM-like component.
We note that despite the small relative increase, the Match-Tensor\&BM25 model is more accurate than all other ensemble variants and is nearly 1\% more accurate than Match-Tensor(Exact only)+SSM\&BM25.
Thus, the Match-Tensor model is not only the most accurate model in this list, it also largely subsumes the semantic matching signals in SSM and the relevance matching signals in BM25 as indicated by the relative small improvement in accuracy when BM25 is added to it.

\begin{table*}[]
\centering
\caption{Relative improvement in model accuracy by combining BM25 with original model.  The smallest change in each column is shown in bold to signify which architecture already best incorporates the signal available in BM25.}
\label{table:results}
\begin{tabular}{r|c|ccc|c}
  Model & \textbf{AUC} & \textbf{NDCG@1} & \textbf{NDCG@3} & \textbf{NDCG@10} & \textbf{ERR}  \\
\hline % Neural + BM25 models
SSM \& BM25                          &3.29\% &2.53\% & 2.11\% & 1.33\% & 1.47\%\\
Match-Tensor \& BM25                 &1.64\% &\textbf{0.94\%} & \textbf{0.79\%} & \textbf{0.54\%} & \textbf{0.57\%}  \\
Match-Tensor+SSM \& BM25             &\textbf{1.52\%} &1.05\% & 0.882\% & 0.65\% & \textbf{0.57\%}\\
Match-Tensor(Exact  only)+SSM \& BM25 &2.32\% &1.12\% & 0.877\% & 0.77\% & 0.72\% \\
\end{tabular}
\end{table*}
%absolute numbers
%SSM \& BM25                          &0.784  &0.850 & 0.872 & 0.911 & 0.691  \\
%Match-Tensor \& BM25                 &0.802  &0.863 & 0.883 & 0.918 & 0.697  \\
%Match-Tensor+SSM \& BM25             &0.802  &0.862 & 0.882 & 0.918 & 0.697  \\
%Math-Tensor+SSM (Exact Only) \& BM25 &0.793  &0.856 & 0.877 & 0.915 & 0.694  \\

\subsection{Model Introspection}

% \usepackage{multirow}
\begin{table*}[]
\caption{Illustrative examples highlighting pairs of results that were incorrectly ordered by a method but were correctly ordered by the Match-Tensor model}
\label{table:model_comparison_bm25}
\begin{tabular}{|p{2.75cm}|p{5cm}|p{5cm}|p{3.25cm}|}
  \hline
  \textbf{Query} & \textbf{Irrelevant Result} & \textbf{Relevant Result} & \textbf{Method with incorrect ranking}\\
  \hline
  ariana tv show & Leah Michele's tv show.. & Ariana on the tv.. & SSM \\
  \hline
  corn shucking song& Blackhawks playing the blues.. & The corn shucking song.. & SSM \\
  \hline
  cats acted like humans& ..humans acted like cats.. & ..cats trying to act like humans.. & BM25\\
  \hline
  low fat high carb& Low carb high fat diet.. & ..popular low fat high carb .. & BM25\\
  \hline
  cleveland wins nba championship& Golden State beats Cleveland in NBA championship.. & Cleveland wins basketball championship.. & Match-Tensor(Exact only) + SSM\\
  \hline
  scholarship to master degree& My score is low for scholarship to master degree .. & ..Master's application and offers scholarship.. & Match-Tensor(Exact only) + SSM\\
  \hline

  \hline
\end{tabular}
\end{table*}

We illustrate the strengths of the Match-Tensor model over other approaches with a few examples in Table \ref{table:model_comparison_bm25}.
SSM is focused towards matching representations. As a result, it often misses out on relevance matching signals by finding a result about the same broad topic but different in several crucial details: as an example, for a query about a celebrity tv show, the model ranks a document about a different celebrity's TV show above a relevant result.

Under its bag of words model, BM25 often scores results that have completely wrong phrases but the right set of tokens above a relevant result. This is best illustrated by the query ``low fat high carb'' where the model prefers a result about ``low carb high fat'' over a relevant result. Traditional learning-to-rank methods address this problem with specifically engineering proximity and ordering features.
Match-Tensor, on the other hand, correctly ranks these results, learning the necessary proximity, ordering, grammatical, and other relationships directly from training examples.

The Match-Tensor(Exact only)+SSM model uses only exact matches between query and document terms and relies on a single representation of query and document to capture semantic signals. This results in subtler failures, often due to an over-reliance on the exact-match channel: for a query inquiring about scholarships for graduate programs, "scholarship to master degree", the exact-only model prefers a document that has the exact phrase but is semantically not useful to the searcher. The full Match-Tensor model correctly prefers another result that matches the intent of the query even though the document doesn't contain an exact match.

%Table.\ref{table:model_comparison_bm25} shows a few illustrative examples highlighting the difference between Match-Tensor and the other models studied here.
%The first example query ``cats act like humans`` illustrates the failure of BM25 to penalize text with the right tokens but wrong order;
%the Match-Tensor correctly ranks this pair of results.
%The second example illustrates the inability of the SSM model to distinguish between semantically similar but distinct entities: for the query "ariana on the tv show", the SSM model scores a document about a different celebrity's tv show higher than a relevant document that doesn't explicitly list the token "show".
%The Match-Tensor model is again able to rank these documents correctly.

%\begin{figure*}
%\includegraphics[width=0.7\textwidth]{bdt}
%\caption{Model performance when adding additional features}
%\label{fig:bdt}
%\end{figure*}

%\begin{table}[]
%\centering
%\caption{NDCG}
%\label{table:ndcg}
%\begin{tabular}{llll}
%\multicolumn{1}{c}{} & \multicolumn{1}{c}{\textbf{NDCG@1}} & \multicolumn{1}{c}{\textbf{NDCG@3}} & \multicolumn{1}{c}{\textbf{NDCG@10}} \\
%BM25 & .758 & .782 & .834 \\
%SSM & .754 & .778 & .830 \\
%MatchTensor & .791 & .810 & .853 \\
%MatchTensor + SSM  & .789 & .809 & .853 \\
%\end{tabular}
%\end{table}

\section{Concluding Discussion}

Deep Neural Networks are a compelling development in machine learning that have substantially advanced the state-of-the-art in several disciplines\cite{lecun2015deep}. While initial developments in several domains were focused on the absolute accuracy~\cite{krizhevsky2012imagenet,sutskever2014sequence} of these models when compared to alternatives, the focus has more recently gravitated towards the completeness of these models; indeed in several domains such as speech recognition, computer vision and machine translation, entire production systems have been completely replaced with neural networks that are trained end-to-end~\cite{hinton2012deep,google_translate}.

Early neural network models for search focused on semantic matching signals which supplemented existing relevance matching features. By computing similarity between semantic representations of the query and document, this class of models naturally captured signals that were hard to determine using traditional models. However, this general class of models appears to miss critical relevance-matching signals \cite{croft_relevance_cikm16}.
In this work, we explored Match-Tensor, a new Deep Relevance model architecture for Search. By simultaneously accounting for several notions of similarity with an expressive 3-D tensor layer and by deferring the combination of these signals into a relevance score to later layers, Match-Tensor is able to achieve higher accuracies than other architectures. More interestingly, this architecture appears to largely subsume the signals in previous models: adding a SSM-like component to the model does not affect the accuracy of the final model, while the improvement when adding BM25 is small and far less than the corresponding improvements in other model architectures.
While we have tailored the details of the model Match-Tensor architecture and the alternatives we studied for the search task within a specific social network, we expect that these learnings might also be adaptable to search within other domains -- just as we have adapted the SSM style models originally developed for web-search to our setting.

The ability to select diverse ways of computing similarity between query and document in the form of channels in the match-tensor model layer is a general and powerful primitive. Indeed, we have only explored a few design choices within this general design space (comparing bi-LSTMs to CNNs). It is possible that increasing the diversity of these sources, using a mix of RNNs, CNNs and other notions of exact matching (by incorporating named entity linking, for example) might further strengthen the accuracy and performance of the model. These, and other explorations are deferred to future work.

% Bibliography
\bibliographystyle{ACM-Reference-Format}
\bibliography{rnn_conv2d}

\newpage
\section*{Appendix: Alternative model architectures with hyperparameters}


\end{document}